\documentclass[12pt]{article}
\usepackage{graphics}
\begin{document}

\large
\begin{enumerate}
\item 14 August 1996 (final accepted version)
\item Non-collinear magnetism
in distorted perovskite compounds
\item I.V.Solovyev$^{a,*}$,
N.Hamada$^b$, K.Terakura$^c$
\item
$^a$JRCAT-ATP, 1-1-4 Higashi, Tsukuba, Ibaraki 305, Japan\\
$^b$Science University of Tokyo, 2641 Yamasaki, Noda, Chiba 278, Japan\\
$^c$JRCAT-NAIR, 1-1-4 Higashi, Tsukuba, Ibaraki 305, Japan\\
\item Abstract \par \hspace{1cm}
Using results of the band structure calculations in the
local-spin-density approximation
we demonstrate how the crystal distortions affect the magnetic
structure of orthorhombically distorted perovskites
leading to a non-collinear spin arrangement.
Our results suggest that the non-collinearity of the spin
magnetic moments, being generally small in La$M$O$_3$
series with $M$=Cr-Fe, is large in SrRuO$_3$.
\item {\it keywords:} Heisenberg exchange,
Dzyaloshinskii-Moriya exchange,
perovskite transition metal oxides,
band structure calculations
\item $^*$Corresponding author.\\
I.V. Solovyev\\
JRCAT-ATP, c/o NAIR\\
1-1-4 Higashi \\
Tsukuba, Ibaraki 305\\
JAPAN\\
Fax:+81-298-54-2788\\ E-mail: igor@jrcat.or.jp
\end{enumerate}


\newpage

  Perovskite transition-metal oxides
are known to be
the most striking example of materials where magnetic, transport and
structural properties are strongly coupled. The reciprocal
influence of the spin and lattice degrees of freedom can be
due to the spin-orbit interaction (SOI) or
the orbital ordering effects \cite{kugel}.
Both mechanisms
can be responsible for the non-collinear magnetic arrangement
through
the
antisymmetric Dzyaloshinskii-Moriya exchange interaction \cite{DM} or
through strong dependence of the interatomic exchange
on the orbital ordering \cite{kugel} resulting in
pronounced non-Heisenberg behavior \cite{nagaev}.
We consider the first possibility for several
orthorhombically distorted perovskites with
$D^{16}_{2h}$ structure
where the non-collinear magnetism is allowed by symmetry
\cite{treves}.
However,
the magnitude of the effect itself depends on the relative strength of
several magnetic interactions.

 We use the LMTO Green's function technique in the real space and
perturbative approach both for small deviations of the spin magnetization
near the scalar-relativistic equilibrium and SOI.
Then, the total energy change
can be expressed analytically as
$\delta E= E_H + E_{DM} + E_{MAE}$.
The first term
$E_H \simeq - 1/2 \sum_{\it ij} J_{\it ij}{\bf e}_{\it i}{\bf e}_{\it j}$
(${\bf e}_{\it i}$ is the direction of the spin magnetization at
the site ${\it i}$)
describes the isotropic Heisenberg exchange interaction and appears
in the second order with respect to nonuniform
rotations of spins
\cite{liecht}. The
magnetocrystalline anisotropy energy (MAE) firstly
appears in the second order with respect to the SOI.
The antisymmetric coupling
$E_{DM} \simeq \sum_{\it i>j} {\bf d}_{\it ij}
[{\bf e}_{\it i} \times {\bf e}_{\it j}]$ corresponds to the
mixed type perturbation with respect to spin rotations and SOI
\cite{igor2}.

  $J_{\it ij}$ and ${\bf d}_{\it ij}$ parameters
for La$M$O$_3$ series with $M$=Cr-Fe and
SrRuO$_3$ are shown in Tables \ref{tab.J_ij} and \ref{tab.d_ij}.
General tendencies of the nearest neighbor interactions $J_{\it ij}$
can be understood by using simple tight-binding arguments given in
\cite{heine}. (i) $J_{\it ij}<0$ at the half of the band filling:
$t_{2g}$-type exchange interaction in LaCrO$_3$ (formal atomic
configuration of Cr is $t_{2g}^3e_g^0$),
both $t_{2g}$ and $e_g$ interactions in LaFeO$_3$ ($t_{2g}^3e_g^2$).
(ii) $J_{\it ij}>0$ at the beginning and at the end of the band
filling: $e_g$ exchange interaction dominating in LaMnO$_3$
($t_{2g}^3e_g^1$) \cite{igor2}. (iii) $J_{\it ij} \sim 0$
around 1/3 and 2/3 of the band filling: SrRuO$_3$ case
($t_{2g}^4e_g^0$).
On the other hand, ${\bf d}_{\it ij}$ parameters being
proportional to the SOI are generally larger in SrRuO$_3$.
The structural factor defined by rotations of the $M$O$_6$
octahedra relative to each other is of the same order of
magnitude for all compounds considered here.
Thus, for
La$M$O$_3$ compounds the Heisenberg exchange
interaction is clearly
the strongest, whereas for SrRuO$_3$ it is considerably smaller
and comparable with the antisymmetric exchange.

  It is particularly interesting in LaMnO$_3$ that
the interlayer exchange coupling
$J_{1B} = \sum_{{\it j} \in B} J_{1 {\it j}}$
with ${\it j}$ running over Mn atoms in the plane ${\it B}$
in Fig.1
crucially
depends on the
Jahn-Teller distortion (JTD)
and
varies between
ferro- (FM) and antiferromagnetic (AFM) \cite{igor2}.
For the pure compound,
the AFM interlayer coupling stabilized by JTD is large enough
($J_{1B} \simeq -1.4$mRy \cite{igor2})
to overcome the antisymmetric interactions and the magnetic
spin structure is
nearly collinear.
Weak FM canting due to
the ${\bf d}_{12}$ interaction
and estimated as
$\sin^{-1}|\alpha_{\bf c}/J_{1B}|$ is less than $2^\circ$
(see \cite{igor2} for details).
However, JTD is suppressed rapidly
with the hole doping,
directly
affecting the interlayer coupling constant $J_{1B}$. At certain
concentration of the holes one expects
$|J_{1B}|\sim|\alpha_{\bf c}|$ and large
non-collinearity.
This tendency qualitatively explains the appearance of the
spin-canted AFM phase
accompanying the AFM-to-FM transition
in the
low-doped manganites \cite{kawano}.

  In conclusion, the non-collinear spin structure, imposed by
general symmetry rules, is suppressed in La$M$O$_3$ oxides
by the strong isotropic exchange interaction. The latter is reduced in
SrRuO$_3$ suggesting essentially non-collinear magnetic arrangement.

  The work
is partly supported by New Energy and Industrial
Technology Development Organization (NEDO).

\newpage

\begin{table}
\caption{Parameters of the isotropic exchange interaction
         $J_{\it ij}$ (in mRy).
         Atomic positions 1, 2 and 3 are shown in Fig.1.}
\label{tab.J_ij}
\begin{center}
\begin{tabular}{rrr}\hline
compound  & $J_{12}$  & $J_{13}$  \\\hline
LaCrO$_3$ & -1.321  & -1.372   \\
LaMnO$_3$ & ~0.225  & ~0.668   \\
LaFeO$_3$ & -2.559  & -3.119   \\
SrRuO$_3$ & ~0.306  & -0.101   \\\hline
\end{tabular}
\end{center}
\end{table}

\begin{table}
\caption{Parameters of the antisymmetric exchange interaction
         ${\bf d}_{\it ij}$ for two $M-$O$-M$ bonds (in mRy).
         The symmetry of the nearest neighbor interactions
         is shown in Fig.1.}
\label{tab.d_ij}
\begin{center}
\begin{tabular}{rcc}\hline
compound  & ${\bf d}_{12}$:(-$\alpha_{\bf c}$,-$\beta_{\bf c}$,~0)  &
${\bf d}_{13}$:(~$\alpha_{\bf ab}$,-$\beta_{\bf ab}$,$\gamma_{\bf ab}$)  \\\hline
LaCrO$_3$ & (-0.005,-0.044,~0) & (~0.029,-0.028,~0.035) \\
LaMnO$_3$ & (-0.032,-0.052,~0) & (~0.032,-0.024,~0.039) \\
LaFeO$_3$ & (-0.019,-0.125,~0) & (~0.075,-0.059,~0.086) \\
SrRuO$_3$ & (-0.138,-0.286,~0) & (-0.062,-0.127,~0.198) \\\hline
\end{tabular}
\end{center}
\end{table}

\newpage

\begin{figure}
\centering \noindent
\resizebox{10cm}{!}{\includegraphics{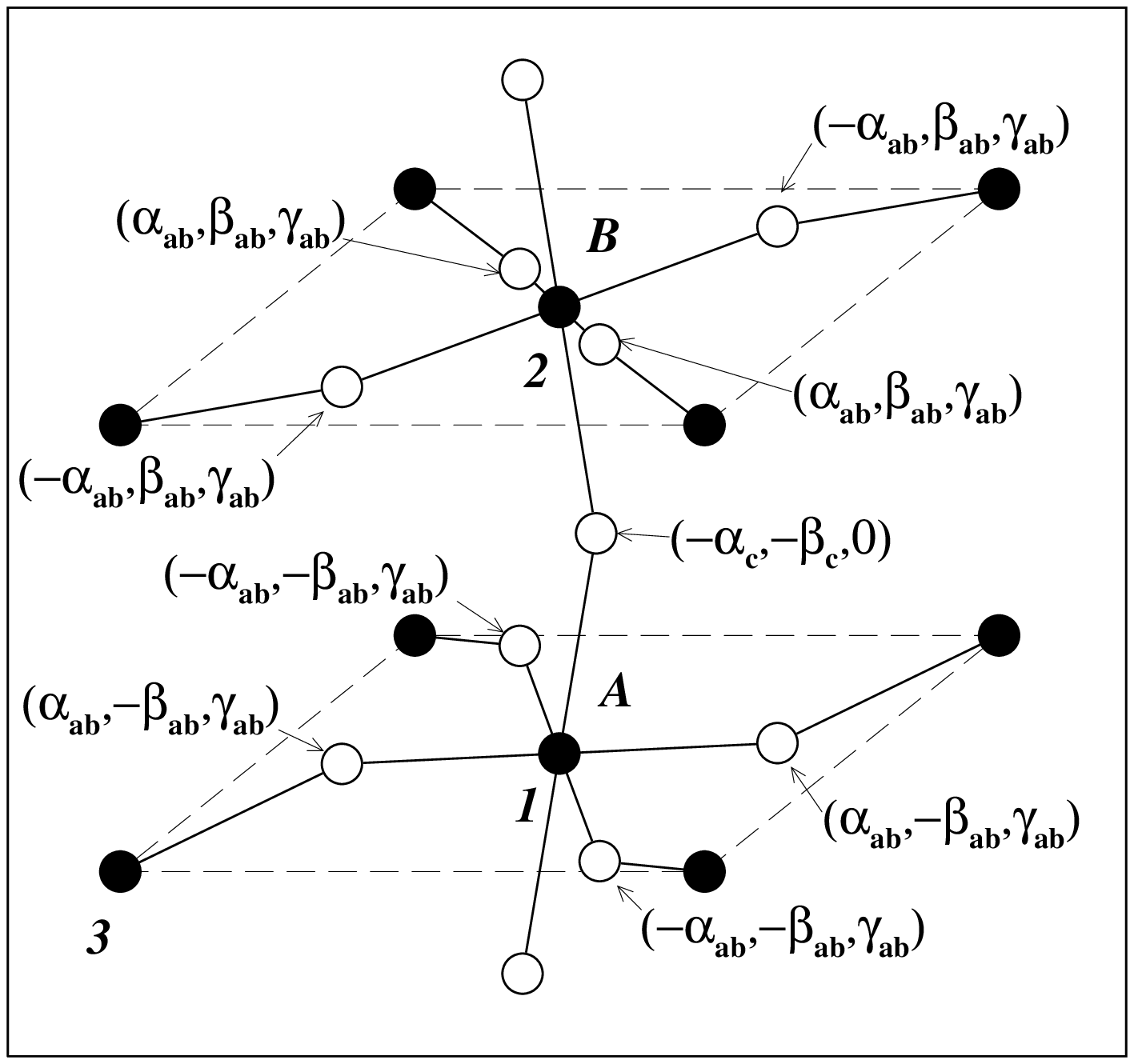}}
\caption{Parameters of the antisymmetric exchange interaction
           associated with different
           $M-$O$-M$ bonds in $D^{16}_{2h}$ structure
           (black and white spheres are $M$ and O
           respectively). $\alpha$, $\beta$ and $\gamma$ are the
           components of ${\bf d}_{\it ij}$ vectors along
           orthorhombic ${\bf a}$,
           ${\bf b}$ and ${\bf c}$ axes for inplane (${\bf ab}$)
           and interplane (${\bf c}$) interactions.}
\label{fig.d_ij}
\end{figure}


\newpage

\begin{thebibliography}{10}

\bibitem{kugel} K.I.Kugel and D.I.Khomskii,
Sov. Phys. Usp. {\bf 25}, 231 (1982).

\bibitem{DM} I.Dzyaloshinskii, J. Phys. Chem. Solids
{\bf 4}, 241 (1958);
T.Moriya, Phys. Rev. {\bf 120}, 91 (1960).

\bibitem{nagaev} E.L.Nagaev,
Sov. Phys. Usp. {\bf 25}, 31 (1982).

\bibitem{treves}
D.Treves, Phys. Rev. {\bf 125}, 1843 (1962).

\bibitem{liecht}
A.I.Liechtenstein {\it et~al.},
J. Magn. Magn. Mater. {\bf 67}, 65 (1987).

\bibitem{igor2}
I.Solovyev {\it et~al.},
Phys. Rev. Lett. {\bf 76}, 4825 (1996).

\bibitem{heine} V.Heine and J.H.Samson, J. Phys. F: Metal Phys.
{\bf 13}, 2155 (1983).

\bibitem{kawano} H.Kawano {\it et~al.},
Phys. Rev. B {\bf 53}, 2202 (1996).

\end{thebibliography}
\end{document}